\documentclass{JHEP3}
\usepackage{graphicx,amsmath,amssymb}
\usepackage{epsfig,multicol}
\usepackage{epsf}
\usepackage{epstopdf}
\input{epsf.sty}

\usepackage{graphicx}
\usepackage{amssymb}
\usepackage{amsmath}

\def\half{{1\over 2}}
\def\({\left(}
\def\){\right)}
\def\[{\left[}
\def\]{\right]}

\def\cn{{\cal N}}

\def\e{\begin{equation}}
\def\q{\end{equation}}
\def\m{\begin{eqnarray}}
\def\n{\end{eqnarray}}

\title{Curvaton with Polynomial Potential}
\author{Qing-Guo Huang \footnote{huangqg@kias.re.kr}
\\\small{\em School of Physics, Korea Institute for Advanced Study,
207-43, Cheongryangri-Dong, Dongdaemun-Gu, Seoul 130-722, Korea } }

\abstract{In general a weakly self-interacting curvaton field is
expected and the curvaton potential takes the polynomial form. The
curvaton potential can be dominated by the self-interaction term
during the period of inflation if the curvaton field stays at a
large vacuum expectation value. We use the $\delta {\cal N}$
formalism to calculate the primordial curvature perturbation in the
various possible scenarios which make the curvaton model much
richer. }


\begin{document}

\section{Introduction}

Inflation \cite{Guth:1980zm,Linde:1981mu,Albrecht:1982wi} is the
leading paradigm to solve the puzzles in the hot big bang model. The
quasi-exponential expansion during inflation makes our universe
almost homogeneous and isotropic. A bonus of inflation is that the
quantum fluctuations of the scalar fields during inflation can
naturally explain the small temperature fluctuations in the cosmic
microwave background radiations and seed the formation of the
large-scale structure. At the leading order, these quantum
fluctuations are characterized by their amplitudes and tilts. If we
only focus on the primordial power spectrum, we cannot distinguish
inflation from curvaton model
\cite{Linde:1996gt,Enqvist:2001zp,Lyth:2001nq,Moroi:2001ct}.

The gravitational dynamics itself introduces important
non-linearities, which will contribute to the final non-Gaussianity
in the large-scale CMB anisotropies. In fact, the CMB
non-Gaussianity \cite{Komatsu:2001rj,Komatsu:2000vy,Komatsu:2002db}
opens a windows to probe the physics of the early universe. A
well-understood ansatz of non-Gaussianity has a local shape. This
kind of non-Gaussianity can be characterized by some non-linearity
parameters $f_{NL}$, $g_{NL}$ and so on: \e \zeta({\bf
x})=\zeta_g({\bf x})+{3\over 5}f_{NL}\zeta_g^2({\bf x})+{9\over
25}g_{NL}\zeta_g^3({\bf x})+..., \label{ztg}\q where $\zeta_g$ is
the linear, Gaussian part of curvature perturbation. The current
bound from WMAP 5yr data \cite{Komatsu:2008hk} is
$-9<f_{NL}^{local}<111$ at $95\%$ CL. Even though a Gaussian
distribution is still consistent with the present experiments, much
of the allowed region for $f_{NL}^{local}<0$ from WMAP 3yr data was
cut. In the single field inflation model $f_{NL}^{local}\sim {\cal
O}(n_s-1)$ \cite{Maldacena:2002vr}, which is constrained by WMAP
($n_s=0.960_{-0.013}^{+0.014}$) \cite{Komatsu:2008hk} to be much
less than unity. However the curvaton model can easily generate a
large local-type non-Gaussianity
\cite{Linde:1996gt,Enqvist:2001zp,Lyth:2001nq,Moroi:2001ct,Lyth:2002my,Bartolo:2003jx,Malik:2006pm,Sasaki:2006kq}.
See \cite{Bartolo:2004if} for a nice review and see
\cite{Huang:2008ze,Ichikawa:2008iq,Multamaki:2008yv,Suyama:2008nt,Beltran:2008ei,Li:2008jn,Li:2008fm,Huang:2008qf,Huang:2008rj,Li:2008tw,Enqvist:2008gk,Huang:2008bg}
for the recent relevant discussions.

In general we can expect that the curvaton field does not have only
a mass term in its potential. Instead of the simplest curvaton
potential $\half m^2\sigma^2$, we adopt a form for the potential
which allows a range of possibilities \e V(\sigma)=\half
m^2\sigma^2+\sum_{n\geq 4} \lambda_n{\sigma^n\over M^{n-4}}. \q The
term with $n>4$ is non-renormalizable and suppressed by a UV scale
$M$. If all of the interaction terms are negligible,
${\delta\rho_\sigma\over \rho_\sigma}\sim 2{\delta \sigma\over
\sigma}+({\delta\sigma\over \sigma})^2$ and thus the second or
higher order non-Gaussianity parameters $(g_{NL},...)$ will be 0. In
the literatures the potential of curvaton is assumed to be dominated
by the mass term. If a subdominant interaction term is taken into
account, the non-linear evolution on large scales is possible. The
curvaton dynamics after inflation was discussed in
\cite{Enqvist:2008gk,Huang:2008bg,Lyth:2003dt,Dimopoulos:2003ss,Enqvist:2005pg}.
In this case $f_{NL}$ can be small even when $f_D\ll 1$, but
$g_{NL}$ should be large
\cite{Sasaki:2006kq,Enqvist:2008gk,Huang:2008bg}. In all of these
papers, the authors only focused on the case where the curvaton
potential is always dominated by the mass term and the interaction
term is taken as a perturbation. However, the self-interaction term
can be dominant if the curvaton mass is small enough and the vacuum
expectation value of curvaton during inflation is large enough. If
so, the higher order non-Gaussianity parameters are also expected to
be larger.

In this paper we will use $\delta {\cal N}$ formalism
\cite{Starobinsky:1986fxa,Sasaki:1995aw,Lyth:2005fi} to calculate
the primordial curvature perturbation for the curvaton model with a
polynomial potential. When the self-interaction term is taken into
account, the curvaton model becomes much richer. Our paper is
organized as follows. In Sec. 2, we calculate the primordial power
spectrum and the non-linearity parameters in various possible
scenarios. In Sec. 3, the spectral index of the primordial power
spectrum and the enhancement of the second order non-Gaussianity
parameters are discussed. The evolution of curvaton before it starts
to oscillate, and after it starts to oscillate, but before it
decays, are investigated in Sec. 4 and Sec. 5 respectively. In Sec.
6, we give some discussions on curvaton model.

\section{Primordial curvature perturbation}

In this paper we expand any field or perturbation at each order
$(n)$ as follows \e \zeta(t,{\bf x})=\zeta^{(1)}(t,{\bf
x})+\sum_{n=2}^{\infty}{1\over n!}\zeta^{(n)}(t,{\bf x}). \q We
assume that the first-order term $\zeta^{(1)}$ is Gaussian and
higher-order terms describe the non-Gaussianity of the full
nonlinear $\zeta$. Working in the framework of Fourier
transformation of $\zeta$, the primordial power spectrum ${\cal
P}_\zeta$ is defined by \e \langle\zeta({\bf k_1})\zeta({\bf
k_2})\rangle=(2\pi)^3 {\cal P}_{\zeta}(k_1)\delta^3({\bf k_1}+{\bf
k_2}), \q  and the primordial bispectrum and trispectrum are defined
by \m\langle\zeta({\bf k_1})\zeta({\bf k_2})\zeta({\bf
k_3})\rangle&=&(2\pi)^3 B_\zeta(k_1,k_2,k_3)\delta^3({\bf k_1}+{\bf k_2}+{\bf k_3}), \\
\langle\zeta({\bf k_1})\zeta({\bf k_2})\zeta({\bf k_3})\zeta({\bf
k_4})\rangle&=&(2\pi)^3 T_\zeta(k_1,k_2,k_3,k_4)\delta^3({\bf
k_1}+{\bf k_2}+{\bf k_3}+{\bf k_4}). \n The bispectrum and
trispectrum are respectively related to the power spectrum by \m
B_\zeta(k_1,k_2,k_3)&=&{6\over 5}
f_{NL}[{\cal P}_\zeta(k_1){\cal P}_\zeta(k_2)+2\ \hbox{perms}], \\
T_\zeta(k_1,k_2,k_3,k_4)&=&\tau_{NL}[{\cal P}_\zeta(k_{13}){\cal
P}_\zeta(k_3){\cal P}_\zeta(k_4)+11\ \hbox{perms}] \nonumber \\
&+&{54\over 25}g_{NL}[{\cal P}_\zeta(k_2){\cal P}_\zeta(k_3){\cal
P}_\zeta(k_4)+3\ \hbox{perms}]. \n Here the non-linearity parameter
$\tau_{NL}$ is not an independent non-linearity parameter and it is
given by \e \tau_{NL}={36\over 25}f_{NL}^2. \label{cvtf}\q But
$g_{NL}$ is an independent parameter which will be calculated in
this paper.

The primordial density perturbation can be described in terms of the
nonlinear curvature perturbation on uniform density hypersurfaces
\cite{Lyth:2004gb} \e \zeta(t,{\bf x})=\delta {\cal N} (t,{\bf x})+
{1\over 3}\int _{{\bar \rho}(t)}^{\rho(t,{\bf x})}{d{\tilde
\rho}\over {\tilde \rho}+{\tilde p}}, \label{dln}\q where $\cn=\int
Hdt$ is the integrated local expansion, ${\bar \rho}$ is the
homogeneous density in the background model, ${\tilde \rho}$ is the
local density and ${\tilde p}$ is the local pressure.

For simplicity, the potential of curvaton field is assumed to
contain a mass term and a self-interaction term as follows \e
V(\sigma)=\half m^2\sigma^2+{1 \over n(n-1)} \lambda \sigma^n.
\label{pt}\q The coupling constant $\lambda$ takes dimensions of
$E^{4-n}$. The effective mass of curvaton is given by \e {\tilde
m}=\sqrt{m^2+\lambda \sigma^{n-2}}.\q The potential is dominated by
the interaction term if \e \sigma>\sigma_c=\({n(n-1)m^2\over
2\lambda}\)^{1\over n-2}. \q For a weakly coupled field, the quantum
fluctuations can be well described by a Gaussian random field
\cite{Seery:2005gb}. Since we are also interested in the case where
the self-interaction term dominates the potential, we want to
estimate the effects of the non-linear quantum fluctuations in
curvaton field at Hubble exit during inflation. The fluctuation of
curvaton is expanded to the third order as follows \e \sigma_*={\bar
\sigma}_*+\delta \sigma_*^{(1)}+\half \delta\sigma_*^{(2)}+{1\over
6} \delta\sigma_*^{(3)}, \q where * denotes that the quantities are
evaluated at the Hubble exit during inflation. The perturbations of
a self-interacting scalar field during inflation are discussed in
the Appendix A. The second and the third order perturbations are
respectively related to $\delta\sigma_*^{(1)}$ by Eqs.\eqref{ds2}
and \eqref{ds3}. Here the curvaton potential deviates from the
exactly quadratic form and then the non-linear evolution of curvaton
field on large scales is expected. Generally the initial amplitude
of curvaton oscillations is some function of the field value at the
Hubble exit: \e \sigma_o=\sigma_o(\sigma_*). \q Thus we can expand
$\sigma_o$ around ${\bar \sigma}_o=\sigma_o(\sigma_*)$ as follows \e
\sigma_o={\bar \sigma}_o\[1+X+\half (h_2+\kappa_2)X^2+{1\over
6}(h_3+3h_2\kappa_2+\kappa_3)X^3\], \label{dso}\q where  \m
X&=&{\delta \sigma_o^{(1)}\over
{\bar \sigma_o}},  \\
h_2&=&{{\bar \sigma_o}\sigma_o''\over {\sigma_o'}^2}, \quad
\kappa_2=-(n-2){N_k\over 3H_*^2}{\lambda{\bar \sigma}_o^{n-2}\over \sigma_o'}, \\
h_3&=&{{\bar \sigma}_o^2\sigma_o'''\over {\sigma_o'}^3}, \quad
\kappa_3=-(n-2)(n-3){N_k\over 3H_*^2}{\lambda{\bar
\sigma}_o^{n-2}\over {\sigma_o'}^2}.\n Here the prime denotes the
derivative with respective to $\sigma_*$. Usually $\sigma_o'\sim
{\cal O}(1)$, and then $-\kappa_2\sim -\kappa_3\sim N_k {\lambda
{\bar \sigma}^{n-2}/ H_*^2}\lesssim N_k{\tilde m}^2/H_*^2\ll 1$. So
it is also reasonable to consider that the quantum fluctuations of
curvaton at Hubble exit during inflation can be well described by a
Gaussian random field, namely $\sigma_*={\bar \sigma}_*+\delta
\sigma_*$. The higher order terms can be neglected even when the
curvaton self-interaction term dominates its potential. In the
following discussions, we will ignore all of the terms with
$\kappa_2$ and $\kappa_3$. These terms can be recovered by
$h_2\rightarrow h_2+\kappa_2$ and $h_3\rightarrow
h_3+3h_2\kappa_2+\kappa_3$ if one want.

In order to make our calculations clearer, we calculate the
curvature perturbation in the various possible scenarios separately.

\subsection{Curvaton potential is dominated by the mass term during inflation}

In this case, the value of curvaton during inflation satisfies
$\sigma_*\ll \sigma_c$ and the interaction term can be taken as a
perturbation. The curvature perturbation in this case has been
discussed very well. To make our paper complete, we directly quote
the results from \cite{Sasaki:2006kq}.

The amplitude of the primordial power spectrum and the non-linearity
parameters are respectively given by \m
P_\zeta&=&{1\over 9\pi^2}f_D^2q^2{H_*^2\over \sigma_*^2}, \\
f_{NL}&=&{5\over 4f_D}(1+h_2)-{5\over 3}-{5f_D\over 6}, \\
g_{NL}&=& {25\over 54}\[{9\over 4f_D^2}(h_3+3h_2)-{9\over
f_D}(1+h_2)+\half (1-9h_2)+10f_D+3f_D^2 \right], \n where \e
q={\sigma_*\sigma_o'\over \sigma_o}, \quad
f_D={3\Omega_{\sigma,D}\over 4-\Omega_{\sigma,D}},\q where
$\Omega_{\sigma,D}$ is the fraction of curvaton energy density in
the energy budget at the time of curvaton decay. If the curvaton
potential is purely quadratic, $h_2=h_3=0$ and then \e
g_{NL}+{10\over 3}f_{NL}\simeq 0. \q Any deviation from the above
relation implies that the curvaton potential does not take the
purely quadratic form.

\subsection{Curvaton potential is dominated by the interaction term during inflation}

In this subsection, we focus on the cases in which the
self-interaction term dominates the curvaton potential during
inflation. The value of curvaton is roughly the same as that when it
starts to oscillate. So we also assume that the curvaton energy
density is dominated by the self-interaction term when it starts to
oscillate. According to Eq.\eqref{dso}, the curvaton density
fluctuation during the curvaton oscillation can be expanded as \e
\rho_{\sigma_o}={\bar \rho}_{\sigma_o}\left(1+nX+{n\over
2}(n-1+h_2)X^2+{n\over 6}[(n-1)(n-2)+3(n-1)h_2+h_3]X^3\right).
\label{rrc}\q As demonstrated in \cite{Turner:1983he}, the energy
density of an oscillating scalar field in an expanding universe with
potential $V\sim \sigma^n$ scales as \e \rho_\sigma\sim
a^{-6n/(n+2)}. \q Or equivalently, the pressure of curvaton when it
is oscillating is related to its energy density by \e p={n-2\over
n+2}\rho. \q When curvaton starts to oscillate, but before it
decays, the non-linear curvature perturbation on uniform-curvaton
density hypersurfaces is given by \e \zeta_{\sigma_o}(t,{\bf
x})=\delta {\cal N}(t,{\bf x})+{n+2\over 6n}\int_{{\bar
\rho}_{\sigma_o}}^{\rho_{\sigma_o}(t,{\bf x})}{d{\tilde
\rho}_{\sigma_o}\over {\tilde \rho}_{\sigma_o}}.\q Therefore the
curvaton density on spatially flat hypersurfaces is \e
\rho_{\sigma_o}|_{\delta{\cal N}=0}=\exp\[{6n\over
n+2}\zeta_{\sigma_o}\]{\bar \rho}_{\sigma_o}. \label{cvo}\q
Considering Eq.\eqref{rrc}, order by order, we
obtain \m \zeta_{\sigma_o}^{(1)}&=&{n+2\over 6}X, \\
\zeta_{\sigma_o}^{(2)}&=&-{6\over
n+2}(1-h_2)\(\zeta_{\sigma_o}^{(1)}\)^2, \\
\zeta_{\sigma_o}^{(3)}&=&\({6\over n+2}\)^2 (2-3h_2+h_3)
\(\zeta_{\sigma_o}^{(1)}\)^3.\n

The energy density of the oscillating curvaton decreases as
$\rho_\sigma\sim a^{-6n/(n+2)}$. If $n<4$, the energy density of
curvaton increases with respect to radiation, but decreases with
respect to radiation if $n>4$. On the other hand, the amplitude of
the curvaton oscillations also decreases, and it is possible that
the self-interaction term becomes subdominant before it decays. But
it is also possible that this transition does not happen before
curvaton decays. We will investigate these two possibilities in Sec.
5 in detail. Here we calculate the primordial curvature perturbation
for these two cases respectively.

\subsubsection{Curvaton potential is dominated by the interaction
term before it decays}

The curvaton-decay hypersurface is a uniform-density hypersurface
and thus from Eq.(\ref{dln}) the perturbed expansion on this
hypersurface is $\delta \cn=\zeta$, where $\zeta$ is the total
curvature perturbation at curvaton-decay hypersurface. Before the
curvaton decays, there have been radiations which are the
productions of inflaton decay. Since the equation of state of
radiations is $p_r=\rho_r/3$, the curvature perturbation related to
radiations is \e \zeta_r=\zeta+{1\over 4}\ln{\rho_r\over {\bar
\rho}_r}. \q The pressure of the oscillating curvaton is
$p={n-2\over n+2}\rho$ and thus \e \zeta_{\sigma_o}=\zeta+{n+2\over
6n}\ln{\rho_{\sigma_o}\over {\bar \rho}_{\sigma_o}}. \q In the
absence of interactions between radiations and curvaton, the
curvature perturbations $\zeta_r$ and $\zeta_{\sigma_o}$ are
conserved respectively and
the above two equations can be written as \m \rho_r&=&{\bar \rho}_r\exp\[4(\zeta_r-\zeta) \], \\
\rho_{\sigma_o}&=&{\bar \rho}_{\sigma_o}\exp\[{6n\over
n+2}(\zeta_{\sigma_o}-\zeta)\].\n At the time of curvaton decay, the
total energy density $\rho_{tot}$ is conserved, i.e. \e
\rho_r(t_D,{\bf x})+\rho_{\sigma_o}(t_D,{\bf x})={\bar
\rho}_{tot}(t_D). \q Requiring that the total energy density is
uniform on the decay surface, we have \e
\(1-\Omega_{\sigma,D}\)e^{4(\zeta_r-\zeta)}+
\Omega_{\sigma,D}e^{{6n\over n+2}(\zeta_{\sigma_o}-\zeta)}=1,
\label{nlc}\q where $\Omega_{\sigma,D}={\bar \rho}_{\sigma,D}/{\bar
\rho}_{tot}$ is the fraction of curvaton energy density in the
energy budget at the time of curvaton decay. Here we assume the
curvaton suddenly decays into radiation. In curvaton model, usually
we also assume the curvature perturbation generated by inflaton is
very small and can be ignored, e.g. $\zeta_r=0$. Order by order from
Eq.\eqref{nlc}, we have \m \zeta^{(1)}&=&f_D\zeta_{\sigma_o}^{(1)}, \\
\zeta^{(2)}&=&\[{6(n-1+h_2)\over (n+2)f_D}-{8(n-1)\over
n+2}-{2(4-n)\over n+2}f_D\]\(\zeta^{(1)}\)^2, \\
\zeta^{(3)}&=&\[{36\over (n+2)^2f_D^2}[(n-1)(n-2)+h_3+3(n-1)h_2]
\right. \nonumber \\ &-&\left. {144\over
(n+2)^2f_D}(n-1)(n-1+h_2)+{4\over
(n+2)^2}[44n^2-121n+68-9(4-n)h_2]\right. \nonumber \\
&-&\left. {80\over (n+2)^2}(n-1)(n-4)f_D+{12\over
(n+2)^2}(4-n)^2f_D^2
\] \(\zeta^{(1)}\)^3, \n where \e f_D={3n\Omega_{\sigma,D}\over
2(n+2)-(4-n)\Omega_{\sigma,D}}. \q Therefore the amplitude of the
primordial power spectrum is \e P_\zeta=\({n+2\over
12\pi}\)^2f_D^2q^2{H_*^2\over \sigma_*^2}.\q Identifying
$\zeta^{(1)}=\zeta_g$ and recalling $\zeta=\zeta^{(1)}+\half
\zeta^{(2)}+{1\over 6}\zeta^{(3)}$, from Eq.\eqref{ztg} the
non-linearity parameters are given by \m f_{NL}&=&{5\over
6}\[{6(n-1+h_2)\over (n+2)f_D}-{8(n-1)\over n+2}-{2(4-n)\over
n+2}f_D\],\\ g_{NL}&=&{25\over 54}\[{36\over
(n+2)^2f_D^2}[(n-1)(n-2)+h_3+3(n-1)h_2] \right. \nonumber \\
&-&\left. {144\over (n+2)^2f_D}(n-1)(n-1+h_2)+{4\over
(n+2)^2}[44n^2-121n+68-9(4-n)h_2]\right. \nonumber \\
&-&\left. {80\over (n+2)^2}(n-1)(n-4)f_D+{12\over
(n+2)^2}(4-n)^2f_D^2
\]. \n For $n=2$, these results are just the same as those in Sec.
2.1. For $n\neq 2$, if $f_D\ll 1$, $g_{NL}\simeq
50(n-1)(n-2)/(3(n+2)^2f_D^2)$ which is large, and \e g_{NL}\simeq
{2(n-2)\over 3(n-1)}f_{NL}^2. \q For $n>2$, $g_{NL}$ is positive.

These results can be easily understood. The energy density of
curvaton is $\rho_\sigma \sim \sigma^n$. Considering
$\sigma\rightarrow \sigma+\delta\sigma$, we have \e {\delta
\rho_\sigma\over \rho_\sigma}\sim n{\delta \sigma\over \sigma}+\half
n(n-1)({\delta \sigma\over \sigma})^2+{1\over 6}n(n-1)(n-2)({\delta
\sigma\over \sigma})^3. \label{drr} \q Since $\zeta_g\simeq
{n+2\over 6}f_D{\delta \sigma\over \sigma}$, the curvature
perturbation reads \e \zeta\simeq \zeta_g+{3(n-1)\over
(n+2)f_D}\zeta_g^2+{6(n-1)(n-2)\over (n+2)^2f_D^2}\zeta_g^3. \q
Using Eq.\eqref{ztg}, we find $f_{NL}\simeq  {5(n-1)\over
(n+2)}{1\over f_D}$ and $g_{NL}\simeq {50(n-1)(n-2)\over
3(n+2)^2}{1\over f_D^2}$.

\subsubsection{The mass term becomes dominant before curvaton decays}

The equation of state of the oscillating curvaton is $p={n-2\over
n+2}\rho$ when $\lambda \sigma^n$ is dominant, and $p=0$ when the
mass term is dominant. In this case, there is a transition from
$p={n-2\over n+2}\rho$ to $p=0$ when the amplitude of the curvaton
oscillations is roughly $\sigma_c$. Since the pressure of an
oscillating curvaton field is a unique function of its energy
density, the energy conservation implies that the the curvature
perturbation $\zeta_{\sigma_o}$ is conserved \cite{Lyth:2003im} even
when the equation of state of the oscillating curvaton changes.

In this case, the pressure of the oscillating curvaton is $p=0$
before it decays and thus \e \zeta_{\sigma_o}=\zeta+{1\over
3}\ln{\rho_{\sigma_o}\over {\bar \rho}_{\sigma_o}}. \q Similarly, on
the curvaton-decay hypersurface, we have \e
(1-\Omega_{\sigma,D})e^{4(\zeta_r-\zeta)}+\Omega_{\sigma,D}e^{3(\zeta_{\sigma_o}-\zeta)}=1.
\q Order by order, the curvature perturbation reads
\m \zeta^{(1)}&=&f_D\zeta_{\sigma_o}^{(1)}, \\
\zeta^{(2)}&=&\[{3(n+2h_2)\over (n+2)f_D}-2-f_D\]\(\zeta^{(1)}\)^2, \\
\zeta^{(3)}&=&\[{9\over (n+2)^2f_D^2}[n(n-2)+4h_3+6nh_2] \right.
\nonumber \\ &-&\left. {18\over (n+2)f_D}(n+2h_2)+{2\over
n+2}(5-2n-9h_2)+10f_D+3f_D^2\]\(\zeta^{(1)}\)^3, \n where \e
f_D={3\Omega_{\sigma,D}\over 4-\Omega_{\sigma,D}}. \q The amplitude
of the primordial power spectrum and the non-linearity parameters
are \m P_\zeta&=&\({n+2\over 12\pi}\)^2f_D^2q^2{H_*^2\over \sigma_*^2}, \\ f_{NL}&=&{5\over 6}\[{3(n+2h_2)\over (n+2)f_D}-2-f_D\], \\
g_{NL}&=&{25\over 54}\[{9\over (n+2)^2f_D^2}[n(n-2)+4h_3+6nh_2]
\right. \nonumber \\ &-&\left. {18\over (n+2)f_D}(n+2h_2)+{2\over
n+2}(5-2n-9h_2)+10f_D+3f_D^2\].\n For $n=2$, these results are also
the same as those in Sec. 2.1. For $n\neq 2$, if $f_D\ll 1$,
$g_{NL}\simeq 25n(n-2)/(6(n+2)^2f_D^2)$ which is large, and \e
g_{NL}\simeq {2(n-2)\over 3n}f_{NL}^2, \q which is different from
that in Sec. 2.2.1.



\section{Spectral index of the primordial power spectrum and a
mixed scenario}

The spectral index is an important quantity to characterize the
primordial power spectrum. In curvaton model, the scale dependence
of the primordial power spectrum is the same as that of
$\delta\sigma^{(1)}$. So the spectral index of the primordial power
spectrum in the curvaton model takes the form \e
n_s^{cv}=1+{2{\tilde m}_*^2\over 3H_*^2}-2\epsilon, \q where
${\tilde m}_*$ is the effective mass of curvaton at
$\sigma=\sigma_*$ and $\epsilon=-{\dot H_*\over H_*^2}$ is a
slow-roll parameter. This result is valid for all of the previous
scenarios. Since we have ${\tilde m}_*\ll H_*$ in the curvaton
model, a small value of $\epsilon$ and a closely scale-invariant
power spectrum are expected. However WMAP 5yr data prefers a
red-tilted power spectrum. In \cite{Huang:2008rj}, we suggested a
mixed scenario in which the curvature perturbation generated by
inflaton also makes a significant contribution to the primordial
power spectrum in order to naturally obtain a red-tilted power
spectrum in curvaton model. Denote the curvature perturbation
generated by curvaton as $P_\zeta^{cv}$. If $P_\zeta^{cv}=\beta
P_\zeta^{tot}$, the spectral index of the total primordial power
spectrum becomes \e n_s=\beta n_s^{cv}+(1-\beta)n_s^{inf}, \q where
$n_s^{inf}=1-6\epsilon+2\eta$ is the spectral index of the power
spectrum generated by inflaton. Now the bispectrum and trispectrum
are respectively related to the power spectrum by \m
B_\zeta(k_1,k_2,k_3)&\simeq&{6\over 5}
f_{NL}^{cv}[{\cal P}_\zeta^{cv}(k_1){\cal P}_\zeta^{cv}(k_2)+2\ \hbox{perms}], \\
T_\zeta(k_1,k_2,k_3,k_4)&\simeq&\tau_{NL}^{cv}[{\cal
P}_\zeta^{cv}(k_{13}){\cal
P}_\zeta^{cv}(k_3){\cal P}_\zeta^{cv}(k_4)+11\ \hbox{perms}] \nonumber \\
&+&{54\over 25}g_{NL}^{cv}[{\cal P}_\zeta^{cv}(k_2){\cal
P}_\zeta^{cv}(k_3){\cal P}_\zeta^{cv}(k_4)+3\ \hbox{perms}], \n
where we ignore the contribution to the non-linearity parameters
from the fluctuation of inflaton. Since ${\cal P}_\zeta^{cv}=\beta
{\cal P}_\zeta^{tot}$, the observed non-Gaussianity parameters
become \e f_{NL}\simeq \beta^2f_{NL}^{cv},\quad \tau_{NL}\simeq
\beta^3\tau_{NL}^{cv}, \quad g_{NL}\simeq \beta^3g_{NL}^{cv}. \q
Considering Eq.(\ref{cvtf}), we have \e \tau_{NL}={36\over
25\beta}(f_{NL})^2. \q In Sec. 2, we conclude that the second order
non-Gaussianity parameter $g_{NL}^{cv}$ is proportional to
$(f_{NL}^{cv})^2$, i.e.  $g_{NL}^{cv}=c (f_{NL}^{cv})^2$ where the
coefficient $c$ is different in different case. Similarly we have \e
g_{NL}={c\over \beta}(f_{NL})^2. \q If $\epsilon\simeq 0$,
$n_s\simeq 1+2(1-\beta)\eta$ and a red tilted primordial power
spectrum is obtained if $\beta< 1$. Now the second order
non-Gaussianity parameters are enhanced by a factor $1/\beta$ for a
fixed $f_{NL}$. Or equivalently, the bound on $g_{NL}$ from
experiments will give a bound on $\beta$ for a given $f_{NL}$.

\section{Curvaton dynamics and the non-linearity parameters}

After inflation our universe is dominated by radiation and the
Hubble parameter goes like $H=1/(2t)$. Usually we assume the
curvaton field does not evolve until the Hubble parameter drops
below the effective mass of curvaton. Once $H\sim {\cal O}({\tilde
m})$, the curvaton starts to oscillate around $\sigma=0$. However
the curvaton field slowly evolves even when $H>{\tilde m}$ and the
non-linear evolution is also expected if the interaction term is
taken into account. The evolution of curvaton after inflation, but
before it oscillates, has been discussed in
\cite{Enqvist:2008gk,Huang:2008bg,Lyth:2003dt,Dimopoulos:2003ss,Enqvist:2005pg}
where the interaction term is regarded as a perturbation.

Here we pay attention to the case where the interaction term is
dominant before the curvaton starts to oscillate. The curvaton
equation of motion after inflation is \e \ddot \sigma+{3\over
2t}\dot \sigma=-{\lambda \over n-1}\sigma^{n-1}. \q It is difficult
to find an analytic solution for this non-linear differential
equation. Before curvaton starts to oscillate, the effective
curvaton mass is smaller than the Hubble parameter. So the curvaton
slowly rolls down its potential. Taking the slow-roll approximation,
the curvaton equation of motion is simplified to be \e {3\over
2t}\dot \sigma\simeq -{\lambda\over n-1}\sigma^{n-1}\q whose
solution with the initial condition $\sigma_{ini}=\sigma_*$ at $t=0$
is given by \e \sigma(t)\simeq \sigma_*-{\lambda t^2\over
3(n-1)}\sigma_*^{n-1}. \q The curvaton begins to oscillate roughly
at the time of $t_o=1/(2{\tilde m}_o)$ which corresponds to
$H={\tilde m}_o\simeq \sqrt{\lambda \sigma_*^{n-2}}$.
 Now the
parameters $q$, $h_2$ and $h_3$ are respectively given by \m
q&\simeq& 1-{x_o^2\over 3}, \\ h_2&\simeq&-(n-2){x_o^2\over 3}, \\
h_3&\simeq& -(n-2)(n-3){x_o^2\over 3}, \n where $x_o={\tilde
m}t_o\simeq 1/2$. For $n=4$, $q\simeq 0.92$, $h_2\simeq h_3\simeq
-0.17$. The corrections to the non-linearity parameters from the
curvaton dynamics after inflation, but before curvaton begins to
oscillate, are at the sub-leading order for the case where the
interaction term is dominant before it starts to oscillate.

\section{Evolution of curvaton after curvaton starts to oscillate}

In this section, we assume our universe is dominated by radiation
before the curvaton decays in order for a large non-Gaussianity. The
case in which the curvaton potential is always dominated by the mass
term has been discussed very well. Here we only focus on the
evolution of the oscillating curvaton whose potential is dominated
by the interaction term when it starts to oscillate.

Assume $a=1$ at the time of curvaton starting to oscillate. At this
time the effective mass of curvaton is \e {\tilde
m}_o=\sqrt{m^2+\lambda\sigma_*^{n-2}}\simeq
\sqrt{\lambda\sigma_*^{n-2}},\q and the energy density of curvaton
is \e \rho_{\sigma_o}\simeq {\lambda \over n(n-1)}\sigma_*^n. \q
Here we assume $\sigma_o=\sigma_*>\sigma_c$. The curvaton energy
density drops as $\rho_\sigma=\rho_{\sigma_o}a^{-6n/(n+2)}$. When
the universe evolves to \e a=a_c=\({\sigma_*\over
\sigma_c}\)^{n+2\over 6}, \q the mass term begins to be dominant.
The curvaton starts to decay at the time when the Hubble parameter
drops below the curvaton decay rate $\Gamma_\sigma$ and the scale
factor is \e a_\Gamma=\sqrt{{\tilde m}_o\over \Gamma_\sigma}. \q If
$a_c<a_\Gamma$, the curvaton potential is dominated by the mass term
before it decays, and then \e \Omega_{\sigma,D}\simeq
{\sigma_*^2\over 6M_p^2}{m^2\over {\tilde m}_o^{3\over 2}
\Gamma_\sigma^\half}\({\sigma_*\over \sigma_c}\)^{{n\over 2}-1}. \q
If $a_c>a_\Gamma$, the curvaton potential is always dominated by the
interaction term before it decays and we have \e
\Omega_{\sigma,D}\simeq {\sigma_*^2\over 3n(n-1)M_p^2}\({{\tilde
m}_o\over \Gamma_\sigma}\)^{4-n\over 2+n}. \q

The curvaton energy density does not decrease until it starts to
oscillate. So the curvaton energy density increases with respect to
radiation before it starts to oscillate for an arbitrary value of
$n$. At the time when the curvaton starts to oscillate $(H={\tilde
m}_o)$, we have \e
\Omega_{\sigma,o}={\lambda\sigma_*^n/(n(n-1))\over
3M_p^2H^2}={\sigma_*^2\over 3n(n-1)M_p^2}, \q which can be ${\cal
O}(1)$ if $\sigma_*\sim M_p$, even though the energy density of
curvaton is negligible during inflation. The energy density of the
oscillating curvaton with $n=4$ goes like $a^{-4}$ which is the same
as radiation, and then $\Omega_{\sigma,D}=\Omega_{\sigma,o}$. For
$n>4$, the energy density of the oscillating curvaton decreased with
respect to radiation and thus $\Omega_{\sigma,D}$ is suppressed by a
factor $\({\Gamma_\sigma}/{\tilde m}_o \)^{n-4\over 2+n}$. One point
we want to stress is that $\Omega_{\sigma,D}$ can be much larger
than the fraction of curvaton energy density in the energy budget
during inflation.

Here we also want to estimate the typical value of curvaton during
inflation. The behavior of a light scalar field in de Sitter space
has been studied in
\cite{Bunch:1978yq,Vilenkin:1982wt,Linde:1982uu,Starobinsky:1994bd}.
The quantum fluctuation can be taken as the random walk: \e
\langle\sigma^2\rangle={H_*^3\over 4\pi^2}t. \q On the other hand,
the long wavelength modes of the light scalar field are in the
slow-roll regime and obey the slow-roll equation of motion, i.e. \e
3H_*{d\sigma \over dt}=-{dV(\sigma)\over
d\sigma}=-m^2\sigma-{\lambda \over n-1}\sigma^{n-1}.\q Combining
these two considerations, we have \e {d\langle\sigma^2\rangle \over
dt}={H_*^3\over 4\pi^2}-{2m^2\over
3H_*}\langle\sigma^2\rangle-{2\lambda \over
3(n-1)H_*}\langle\sigma^2\rangle^{n\over 2}. \q For $n=4$, our
result is the same as that in \cite{Starobinsky:1994bd}. In the case
with a dominant interaction term, the solution of the above
differential equation approaches a constant equilibrium value \e
\sigma_*\simeq \({3(n-1)H_*^4\over 8\pi^2\lambda}\)^{1\over n} \q
which can be estimated as the typical value of the curvaton during
inflation. Now the requirement of $\sigma_*>\sigma_c$ yields \e
H_*^{2n-4}>{3(n-1)\over 8\pi^2}\({4n\pi^2\over 3}\)^{n\over
2}{m^n\over \lambda}.\q As a concrete example, we investigate the
case of $n=4$ in the following subsection.

\subsection{The case of $n=4$ with $\sigma_*>\sigma_c$}

In this subsection, we estimate the value of curvaton during
inflation as the typical value, namely \e \sigma_*\simeq
0.58\lambda^{-1/4}H_*, \q and the effective mass of curvaton when it
starts to oscillate is \e {\tilde m}_o\simeq 0.58\lambda^{1/4}H_*.
\q In this case, $\sigma_c=\sqrt{6}\lambda^{-1/2}m$ and
$\sigma_*>\sigma_c$ says \e m<0.24\lambda^{1/ 4}H_*. \q We also have
\e a_c=0.24\lambda^{1/4}{H_*\over m}, \quad
a_\Gamma=0.76\lambda^{1/8}\sqrt{H_*\over \Gamma_\sigma}.\q Requiring
$a_c<a_\Gamma$ yields $\Gamma_\sigma\lesssim \Gamma_c=
10\lambda^{-1/4}m^2/H_*$. If the curvaton decay rate is roughly the
same as the gravitational strength decay rate $\Gamma_g=m^3/M_p^2$,
$a_\Gamma\sim a_c\sqrt{M_p^2\over mH_*\lambda^{1/4}}>a_c$ and the
mass term becomes dominant before curvaton decays. On the other
hand, the gravitational wave perturbation only depends on the
inflation scale and $H_*$ is related to the tensor-scalar ratio $r$
by $H_*=10^{-4}\sqrt{r}M_p$.
\\

\noindent $\bullet$ $a_c>a_\Gamma$. The interaction term always
dominates the curvaton potential before the curvaton decays. The
amplitude of the primordial power spectrum and the non-linearity
parameter generated by curvaton field are respectively given by
$P_\zeta^{cv}=0.075\sqrt{\lambda}f_D^2$ and $f_{NL}^{cv}\simeq
5/(2f_D)$. WMAP normalization \cite{Komatsu:2008hk} is
$P_{\zeta,wmap}=2.457\times 10^{-9}$. If the total amplitude of the
primordial power spectrum is contributed by the curvaton
fluctuation,  we have $P_\zeta^{cv}=P_{\zeta,wmap}$ which implies
$\sqrt{\lambda}\geq 3.3\times 10^{-8}$ because
$f_D=\Omega_{\sigma,D}\leq 1$. On the other hand, we have
$f_D=\Omega_{\sigma,D}\simeq 9.3\times 10^{-3}\lambda^{-1/2}{H_*^2/
M_p^2}$, and then \e f_{NL}=f_{NL}^{cv}=2.7\times 10^2
{\sqrt{\lambda}M_p^2\over H_*^2}\gtrsim {891\over r}.\q The limit of
$r$ from WMAP5 is $r<0.2$. The non-Gaussianity parameter $f_{NL}$ is
much larger than the upper bound from WMAP 5yr data and the above
scenario has been ruled out. On the other hand, in order to make
this point clearer, let's start with the bound on $f_{NL}^{cv}$.
Requiring
$f_{NL}^{cv}=5/(2\Omega_{\sigma,D})=90M_p^2/\sigma_*^2\lesssim 111$
yields $\sigma_*^2\gtrsim 0.81M_p^2$. Since $f_D\leq 1$,
$P_\zeta^{cv}={1\over 4\pi^2}f_D^2{H_*^2\over \sigma_*^2}\lesssim
3\times 10^{-10}r < P_{\zeta,wmap}$. In order to satisfy the bound
on the non-Gaussianity from WMAP 5yr data, it is natural to assume
that the fluctuation of inflaton makes a significant contribution to
the primordial power spectrum, namely $P_\zeta^{cv}=\beta
P_{\zeta,wmap}$. Now the above constraints are released to be
$\sqrt{\lambda}\geq 3.3\times 10^{-8}\beta$ and \e f_{NL}\geq
{891\beta^3\over r}.\q Considering $f_{NL}<111$ and $r<0.2$ yields
$\beta\lesssim 0.3$ and thus $g_{NL}\gtrsim 3c(f_{NL})^2$. If
$f_{NL}=30$ and $r=10^{-3}$, $\beta\lesssim 0.03$ and $g_{NL}\gtrsim
3\times 10^4c$ which can be detected by Planck.
\\

\noindent $\bullet$ $a_c<a_\Gamma$. We also consider
$P_\zeta^{cv}=\beta P_\zeta^{tot}$. In this case,
$P_\zeta^{cv}=0.075\sqrt{\lambda}f_D^2$, $f_{NL}^{cv}\simeq
5/(3f_D)$, and then
$P_\zeta^{tot}=0.21\beta^3\sqrt{\lambda}/(f_{NL})^2$. WMAP
normalization \cite{Komatsu:2008hk} reads \e \lambda\simeq 1.4\times
10^{-16}(f_{NL})^4/\beta^6. \q On the other hand, $f_D\simeq {3\over
4}\Omega_{\sigma,D}=2.3\times
10^{-2}mH_*^{3/2}/(\lambda^{5/8}M_p^2\Gamma_\sigma^{1/2})$.
Considering $f_{NL}=\beta^2f_{NL}^{cv}$, we have \e f_{NL}\simeq
23\beta^{7/ 6}\sqrt{r}({m^2\over M_p\Gamma_\sigma})^{1/3},
\label{mmf}\q which is compatible with WMAP 5yr data if the curvaton
mass is not too large even when the primordial power spectrum is
generated by curvaton field $(\beta=1)$. In this scenario, we have
$a_c<a_\Gamma$ and $\sigma_*>\sigma_c$, namely \e 2\times
10^{-6}r^{3/4}\sqrt{\Gamma_\sigma M_p}/\beta^{1/4} \leq m \leq
2\times 10^{-22}{r^3\over \beta}{M_p^2\over \Gamma_\sigma}. \q From
Eq.(\ref{mmf}), the lower bound on the curvaton mass is
automatically satisfied for $f_{NL}\gtrsim {\cal O}(1)$. The upper
bound on the curvaton mass leads to a bound on $f_{NL}$ from above
as follow \e f_{NL}\leq 8\times 10^{-14}\beta^\half r^{5\over
2}{M_p\over \Gamma_\sigma}.\label{mgf}\q A large non-Gaussianity is
achieved only when curvaton decay rate is very small compared to
$M_p$. In \cite{Lyth:2003dt} the author pointed out that the
curvaton should decay before neutrino decoupling, namely
$\Gamma_\sigma>\Gamma_0=1.8\times 10^{-43}M_p$. Otherwise the
curvature perturbations may be accompanied by a significant
isocurvature neutrino perturbation. This requirement leads to
$f_{NL}\lesssim 4\times 10^{29}\beta^{1\over 2} r^{5\over 2}$ which
is quite loose. On the other hand, the curvaton model is free from
the constraint of isocurvature perturbation in WMAP 5yr result
\cite{Komatsu:2008hk} if the cold dark matter (CDM) is not the
direct decay product of the curvaton and CDM is generated after the
curvaton decays completely (or equivalently $\Gamma_\sigma\gtrsim
(M_{CDM}/20)^2/M_p$) \cite{Beltran:2008ei}. Combing with
Eq.(\ref{mgf}) yields an upper bound on the mass of CDM \e
M_{CDM}\lesssim 6\times 10^{-6}\beta^{1\over 4}r^{5\over
4}M_p/(f_{NL})^{1/2}. \q In \cite{Lyth:2003dt} another constraint on
$\Gamma_\sigma$ is $\Gamma_\sigma\gtrsim \Gamma_g=m^3/M_p^2$ which
yields \e f_{NL}\lesssim 23\beta^{7/ 6}\sqrt{r}(M_p/m)^{1/3}
\label{mgg}\q directly from Eq.(\ref{mmf}). According to
Eq.(\ref{mgf}), we have \e f_{NL}\lesssim 8\times
10^{-14}\beta^\half r^{5\over 2}{M_p^3\over m^3},\q which is more
restricted than Eq.(\ref{mgg}) if ${m\over M_p}>3.8\times
10^{-6}\beta^{-{1\over 4}}r^{3\over 4}$.
\\

To summarize, if the self-interaction term $\lambda \sigma^4$ is
always dominant, $\Omega_{\sigma,D}$ will be too small and the
non-Gaussianity is too large to fit the WMAP 5yr data unless the
fluctuation of inflaton makes the main contribution to the
primordial power spectrum. The constraint on the model where the
curvaton potential is always dominated by the interaction term with
$n>4$ before curvaton decays will be more stringent. However,
because the energy density of the oscillating curvaton whose
potential is dominated by the mass term grows with respect to the
radiation, the non-Gaussianity can be compatible with WMAP 5yr data
even when the primordial power spectrum is mainly generated by
curvaton.

\section{Discussions}

In this paper we use the $\delta {\cal N}$ formalism to calculate
the primordial curvature perturbation on large scales in the
curvaton model with polynomial potential. The main contribution to
the non-Gaussianity in curvatom model comes from the non-linear
gravitational perturbations, rather than the curvaton
self-interaction. Our calculations are also straightforward to apply
to the case with more complicated curvaton potential. When the
self-interaction term dominates the curvaton potential during
inflation, the order of magnitude of the second order non-linearity
parameters $\tau_{NL}$ and $g_{NL}$ is roughly ${\cal O}(f_{NL}^2)$
if $f_{NL}\gg 1$.

A red-tilted primordial power spectrum can be naturally achieved in
curvaton model if the fluctuation of inflaton also makes a
significant contribution to it \cite{Huang:2008rj}. In this mixed
scenario, it is also possible to detect the non-Gaussianity
generated during inflation in the generalized inflation models
\cite{Chen:2006nt,Bean:2008ga,Li:2008qc,Chen:2007gd,Matsuda:2008hx,Gao:2008dt,Li:2008qv,Xue:2008mk,Li:2008gg}.
For another interesting observation in this mixed scenario, the
second order non-Gaussianity parameters $\tau_{NL}$ and $g_{NL}$ are
enhanced for fixed $f_{NL}$. In addition, multiple curvatons are
generically expected in the fundamental theories, such as string
theory. It is worth investigating the curvature perturbation in
N-vaton \cite{Huang:2008rj} with polynomial potential.


\vspace{1.5cm}

\noindent {\bf Acknowledgments}

\vspace{.5cm}

We would like to thank Sungjay Lee and Yi Wang for useful
discussions.


\newpage

\appendix

\appendix

\section{Perturbations of a light self-interacting scalar field in the inflationary Universe}

In this section we consider the perturbations of a light
self-interacting scalar field $\sigma$ whose energy density is
subdominant in the inflationary Universe. The potential of $\sigma$
is given in Eq.\eqref{pt}.

We expand the curvaton field up to the third order in the
perturbations around the homogeneous background as \e \sigma(t,{\bf
x})=\sigma(t)+\delta\sigma^{(1)}(t,{\bf x})+\half
\delta\sigma^{(2)}(t,{\bf x})+{1\over 6}\delta\sigma^{(3)}(t,{\bf
x}).\q During inflation the equations of motion for the homogeneous
part and the perturbations on large scales in cosmic time are given
by \m \ddot \sigma+3H\dot
\sigma&=&-m^2\sigma-{\lambda \over n-1} \sigma^{n-1}, \\
\ddot {\delta\sigma^{(1)}}+3H\dot {\delta\sigma^{(1)}}&=&-{\tilde
m}^2\delta\sigma^{(1)}, \\ \ddot {\delta\sigma^{(2)}}+3H\dot
{\delta\sigma^{(2)}}&=&-{\tilde m}^2\delta\sigma^{(2)}-(n-2)\lambda
\sigma^{n-3}{\(\delta\sigma^{(1)}\)^2}, \\ \ddot
{\delta\sigma^{(3)}}+3H\dot {\delta\sigma^{(3)}}&=&-{\tilde
m}^2\delta\sigma^{(3)}-3(n-2)\lambda \sigma^{n-3}\delta\sigma^{(1)}\delta\sigma^{(2)} \nonumber \\
&-&(n-2)(n-3)\lambda \sigma^{n-4} \(\delta\sigma^{(1)}\)^3.\n In
slow-roll approximation, \e 3H\dot {\delta\sigma^{(2)}}\simeq
-(n-2)\lambda \sigma^{n-3}{\(\delta\sigma^{(1)}\)^2}, \q whose
solution is roughly given by \cite{Bartolo:2004if,Bernardeau:2003nx}
\e \delta \sigma^{(2)}\sim -{N_k\over 3H^2}(n-2)\lambda
\sigma^{n-3}{\(\delta\sigma^{(1)}\)^2},\label{ds2}\q where
$N_k=\int_{t_k}^{t_{end}}Hdt$ is the number of e-folds between the
end of inflation and the time $t_k$ when the scale of wavenumber $k$
leaves the horizon during inflation. Typically $N_k=60$. Similarly,
the solution of $\delta \sigma^{(3)}$ reads \e
\delta\sigma^{(3)}\sim -{N_k\over 3H^2}(n-2)(n-3)\lambda
\sigma^{n-4}\(\delta\sigma^{(1)}\)^3.\label{ds3}\q Both
$\delta\sigma^{(2)}$ and $\delta\sigma^{(3)}$ are proportional to
the coupling constant. In the limit of $\lambda\rightarrow 0$, we
only need to expand the curvaton field to $\delta\sigma^{(1)}$
without higher order, non-Gaussian terms.



\newpage

\begin{thebibliography}{99}



\bibitem{Guth:1980zm}
  A.~H.~Guth,
  ``The Inflationary Universe: A Possible Solution To The Horizon And Flatness
  Problems,''
  Phys.\ Rev.\  D {\bf 23}, 347 (1981).

\bibitem{Linde:1981mu}
  A.~D.~Linde,
  ``A New Inflationary Universe Scenario: A Possible Solution Of The Horizon,
  Flatness, Homogeneity, Isotropy And Primordial Monopole Problems,''
  Phys.\ Lett.\  B {\bf 108}, 389 (1982).

\bibitem{Albrecht:1982wi}
  A.~Albrecht and P.~J.~Steinhardt,
  ``Cosmology For Grand Unified Theories With Radiatively Induced Symmetry
  Breaking,''
  Phys.\ Rev.\ Lett.\  {\bf 48}, 1220 (1982).



\bibitem{Linde:1996gt}
  A.~D.~Linde and V.~F.~Mukhanov,
  ``Nongaussian isocurvature perturbations from inflation,''
  Phys.\ Rev.\  D {\bf 56}, 535 (1997)
  [arXiv:astro-ph/9610219].

\bibitem{Enqvist:2001zp}
  K.~Enqvist and M.~S.~Sloth,
  ``Adiabatic CMB perturbations in pre big bang string cosmology,''
  Nucl.\ Phys.\  B {\bf 626}, 395 (2002)
  [arXiv:hep-ph/0109214].

\bibitem{Lyth:2001nq}
  D.~H.~Lyth and D.~Wands,
  ``Generating the curvature perturbation without an inflaton,''
  Phys.\ Lett.\  B {\bf 524}, 5 (2002)
  [arXiv:hep-ph/0110002].

\bibitem{Moroi:2001ct}
  T.~Moroi and T.~Takahashi,
  ``Effects of cosmological moduli fields on cosmic microwave background,''
  Phys.\ Lett.\  B {\bf 522}, 215 (2001)
  [Erratum-ibid.\  B {\bf 539}, 303 (2002)]
  [arXiv:hep-ph/0110096].


\bibitem{Komatsu:2001rj}
  E.~Komatsu and D.~N.~Spergel,
  ``Acoustic signatures in the primary microwave background bispectrum,''
  Phys.\ Rev.\  D {\bf 63}, 063002 (2001)
  [arXiv:astro-ph/0005036].

\bibitem{Komatsu:2000vy}
  E.~Komatsu and D.~N.~Spergel,
  ``The cosmic microwave background bispectrum as a test of the physics of
  inflation and probe of the astrophysics of the low-redshift universe,''
  arXiv:astro-ph/0012197.

\bibitem{Komatsu:2002db}
  E.~Komatsu,
  ``The Pursuit of Non-Gaussian Fluctuations in the Cosmic Microwave
  Background,''
  arXiv:astro-ph/0206039.




\bibitem{Komatsu:2008hk}
  E.~Komatsu {\it et al.}  [WMAP Collaboration],
  ``Five-Year Wilkinson Microwave Anisotropy Probe (WMAP)
  Observations:Cosmological Interpretation,''
  arXiv:0803.0547 [astro-ph].

\bibitem{Maldacena:2002vr}
  J.~M.~Maldacena,
  ``Non-Gaussian features of primordial fluctuations in single field
  inflationary models,''
  JHEP {\bf 0305}, 013 (2003)
  [arXiv:astro-ph/0210603].


\bibitem{Lyth:2002my}
  D.~H.~Lyth, C.~Ungarelli and D.~Wands,
  ``The primordial density perturbation in the curvaton scenario,''
  Phys.\ Rev.\  D {\bf 67}, 023503 (2003)
  [arXiv:astro-ph/0208055].

\bibitem{Bartolo:2003jx}
  N.~Bartolo, S.~Matarrese and A.~Riotto,
  ``On non-Gaussianity in the curvaton scenario,''
  Phys.\ Rev.\  D {\bf 69}, 043503 (2004)
  [arXiv:hep-ph/0309033].

\bibitem{Malik:2006pm}
  K.~A.~Malik and D.~H.~Lyth,
  ``A numerical study of non-gaussianity in the curvaton scenario,''
  JCAP {\bf 0609}, 008 (2006)
  [arXiv:astro-ph/0604387].

\bibitem{Sasaki:2006kq}
  M.~Sasaki, J.~Valiviita and D.~Wands,
  ``Non-gaussianity of the primordial perturbation in the curvaton model,''
  Phys.\ Rev.\  D {\bf 74}, 103003 (2006)
  [arXiv:astro-ph/0607627].





\bibitem{Bartolo:2004if}
  N.~Bartolo, E.~Komatsu, S.~Matarrese and A.~Riotto,
  ``Non-Gaussianity from inflation: Theory and observations,''
  Phys.\ Rept.\  {\bf 402}, 103 (2004)
  [arXiv:astro-ph/0406398].


\bibitem{Huang:2008ze}
  Q.~G.~Huang,
  ``Large Non-Gaussianity Implication for Curvaton Scenario,''
  arXiv:0801.0467 [hep-th].

\bibitem{Ichikawa:2008iq}
  K.~Ichikawa, T.~Suyama, T.~Takahashi and M.~Yamaguchi,
  ``Non-Gaussianity, Spectral Index and Tensor Modes in Mixed Inflaton and
  Curvaton Models,''
  Phys.\ Rev.\  D {\bf 78}, 023513 (2008)
  [arXiv:0802.4138 [astro-ph]].

\bibitem{Multamaki:2008yv}
  T.~Multamaki, J.~Sainio and I.~Vilja,
  ``Non-Gaussianity in three fluid curvaton model,''
  arXiv:0803.2637 [astro-ph].

\bibitem{Suyama:2008nt}
  T.~Suyama and F.~Takahashi,
  ``Non-Gaussianity from Symmetry,''
  arXiv:0804.0425 [astro-ph].

\bibitem{Beltran:2008ei}
  M.~Beltran,
  ``Isocurvature, non-gaussianity and the curvaton model,''
  Phys.\ Rev.\  D {\bf 78}, 023530 (2008)
  [arXiv:0804.1097 [astro-ph]].

\bibitem{Li:2008jn}
  M.~Li, C.~Lin, T.~Wang and Y.~Wang,
  ``Non-Gaussianity, Isocurvature Perturbation, Gravitational Waves and a No-Go
  Theorem for Isocurvaton,''
  arXiv:0805.1299 [astro-ph].

\bibitem{Li:2008fm}
  S.~Li, Y.~F.~Cai and Y.~S.~Piao,
  ``DBI-Curvaton,''
  arXiv:0806.2363 [hep-ph].

\bibitem{Huang:2008qf}
  Q.~G.~Huang,
  ``Spectral Index in Curvaton Scenario,''
  Phys.\ Rev.\  D {\bf 78}, 043515 (2008)
  [arXiv:0807.0050 [hep-th]].

\bibitem{Huang:2008rj}
  Q.~G.~Huang,
  ``N-vaton,''
  JCAP {\bf 0809}, 017 (2008)
  [arXiv:0807.1567 [hep-th]].


\bibitem{Li:2008tw}
  M.~Li and C.~Lin,
  ``Reconstruction of the isocurvaton scenario,''
  arXiv:0807.4352 [astro-ph].

\bibitem{Enqvist:2008gk}
  K.~Enqvist and T.~Takahashi,
  ``Signatures of Non-Gaussianity in the Curvaton Model,''
  arXiv:0807.3069 [astro-ph].


\bibitem{Huang:2008bg}
  Q.~G.~Huang and Y.~Wang,
  ``Curvaton Dynamics and the Non-Linearity Parameters in Curvaton Model,''
  arXiv:0808.1168 [hep-th].


\bibitem{Lyth:2003dt}
  D.~H.~Lyth,
  ``Can the curvaton paradigm accommodate a low inflation scale,''
  Phys.\ Lett.\  B {\bf 579}, 239 (2004)
  [arXiv:hep-th/0308110].


\bibitem{Dimopoulos:2003ss}
  K.~Dimopoulos, G.~Lazarides, D.~Lyth and R.~Ruiz de Austri,
  ``Curvaton dynamics,''
  Phys.\ Rev.\  D {\bf 68}, 123515 (2003)
  [arXiv:hep-ph/0308015].


\bibitem{Enqvist:2005pg}
  K.~Enqvist and S.~Nurmi,
  ``Non-gaussianity in curvaton models with nearly quadratic potential,''
  JCAP {\bf 0510}, 013 (2005)
  [arXiv:astro-ph/0508573].


\bibitem{Starobinsky:1986fxa}
  A.~A.~Starobinsky,
  ``Multicomponent de Sitter (Inflationary) Stages and the Generation of
  Perturbations,''
  JETP Lett.\  {\bf 42} (1985) 152.

\bibitem{Sasaki:1995aw}
  M.~Sasaki and E.~D.~Stewart,
  ``A General Analytic Formula For The Spectral Index Of The Density
  Perturbations Produced During Inflation,''
  Prog.\ Theor.\ Phys.\  {\bf 95}, 71 (1996)
  [arXiv:astro-ph/9507001].

\bibitem{Lyth:2005fi}
  D.~H.~Lyth and Y.~Rodriguez,
  ``The inflationary prediction for primordial non-gaussianity,''
  Phys.\ Rev.\ Lett.\  {\bf 95}, 121302 (2005)
  [arXiv:astro-ph/0504045].


\bibitem{Lyth:2004gb}
  D.~H.~Lyth, K.~A.~Malik and M.~Sasaki,
  ``A general proof of the conservation of the curvature perturbation,''
  JCAP {\bf 0505}, 004 (2005)
  [arXiv:astro-ph/0411220].


\bibitem{Seery:2005gb}
  D.~Seery and J.~E.~Lidsey,
  ``Primordial non-gaussianities from multiple-field inflation,''
  JCAP {\bf 0509}, 011 (2005)
  [arXiv:astro-ph/0506056].

\bibitem{Turner:1983he}
  M.~S.~Turner,
  ``Coherent Scalar Field Oscillations In An Expanding Universe,''
  Phys.\ Rev.\  D {\bf 28}, 1243 (1983).

\bibitem{Lyth:2003im}
  D.~H.~Lyth and D.~Wands,
  ``Conserved cosmological perturbations,''
  Phys.\ Rev.\  D {\bf 68}, 103515 (2003)
  [arXiv:astro-ph/0306498].


\bibitem{Bunch:1978yq}
  T.~S.~Bunch and P.~C.~W.~Davies,
  ``Quantum Field Theory In De Sitter Space: Renormalization By Point
  Splitting,''
  Proc.\ Roy.\ Soc.\ Lond.\  A {\bf 360} (1978) 117.


\bibitem{Vilenkin:1982wt}
  A.~Vilenkin and L.~H.~Ford,
  ``Gravitational Effects Upon Cosmological Phase Transitions,''
  Phys.\ Rev.\  D {\bf 26}, 1231 (1982).

\bibitem{Linde:1982uu}
  A.~D.~Linde,
  ``Scalar Field Fluctuations In Expanding Universe And The New Inflationary
  Universe Scenario,''
  Phys.\ Lett.\  B {\bf 116}, 335 (1982).

\bibitem{Starobinsky:1994bd}
  A.~A.~Starobinsky and J.~Yokoyama,
  ``Equilibrium state of a selfinteracting scalar field in the De Sitter
  background,''
  Phys.\ Rev.\  D {\bf 50}, 6357 (1994)
  [arXiv:astro-ph/9407016].



\bibitem{Chen:2006nt}
  X.~Chen, M.~x.~Huang, S.~Kachru and G.~Shiu,
  ``Observational signatures and non-Gaussianities of general single field
  inflation,''
  JCAP {\bf 0701}, 002 (2007)
  [arXiv:hep-th/0605045].

\bibitem{Bean:2008ga}
  R.~Bean, D.~J.~H.~Chung and G.~Geshnizjani,
  ``Reconstructing a general inflationary action,''
  arXiv:0801.0742 [astro-ph].

\bibitem{Li:2008qc}
  M.~Li, T.~Wang and Y.~Wang,
  ``General Single Field Inflation with Large Positive Non-Gaussianity,''
  JCAP {\bf 0803}, 028 (2008)
  [arXiv:0801.0040 [astro-ph]].

\bibitem{Chen:2007gd}
  B.~Chen, Y.~Wang and W.~Xue,
  ``Inflationary NonGaussianity from Thermal Fluctuations,''
  JCAP {\bf 0805}, 014 (2008)
  [arXiv:0712.2345 [hep-th]].

\bibitem{Matsuda:2008hx}
  T.~Matsuda,
  ``Modulated Inflation,''
  arXiv:0801.2648 [hep-ph].

\bibitem{Gao:2008dt}
  X.~Gao,
  ``Primordial Non-Gaussianities of General Multiple Field Inflation,''
  arXiv:0804.1055 [astro-ph].

\bibitem{Li:2008qv}
  S.~W.~Li and W.~Xue,
  ``Revisiting non-Gaussianity of multiple-field inflation from the field
  equation,''
  arXiv:0804.0574 [astro-ph].

\bibitem{Xue:2008mk}
  W.~Xue and B.~Chen,
  ``$\alpha$-vacuum and inflationary bispectrum,''
  arXiv:0806.4109 [hep-th].

\bibitem{Li:2008gg}
  M.~Li and Y.~Wang,
  ``Consistency Relations for Non-Gaussianity,''
  arXiv:0807.3058 [hep-th].




\bibitem{Bernardeau:2003nx}
  F.~Bernardeau, T.~Brunier and J.~P.~Uzan,
  ``High order correlation functions for self interacting scalar field in  de
  Sitter space,''
  Phys.\ Rev.\  D {\bf 69}, 063520 (2004)
  [arXiv:astro-ph/0311422].


\end{thebibliography}
\end{document}